\documentclass{elsart}

\usepackage{graphicx,amssymb}
\usepackage[latin1]{inputenc}
\begin{document}

\begin{frontmatter}

\title{Application of Multifractal Wavelet Analysis to Spontaneous Fermentation Processes}

\author[label1]{V. Ibarra-Junquera}\ead{ vij@ucol.mx},
\author[label2]{J.S. Murgu\'ia},\ead{ondeleto@uaslp.mx}
\author[label1]{P. Escalante-Minakata},\ead{minakata@ipicyt.edu.mx}
\author[label3]{H.C. Rosu}\ead{hcr@ipicyt.edu.mx}

\address[label1]{Faculty of Chemical Sciences, University of Colima, Coquimatl\'an, Col., Mexico}
\address[label2]{DFM-UASLP, San Luis Potos\'{\i}, S.L.P., Mexico}
\address[label3]{Division of Advanced Materials, IPICyT, San Luis Potos\'{\i}, S.L.P., Mexico }

Physica A (2008)\\
{\tiny doi: 10.1016/j.physa.2008.01.083}\\
{\tiny arXiv: 0710.2362 v2; multifr1.tex}

\begin{abstract}

An algorithm is presented here to get more detailed information, of mixed culture type, based exclusively on the biomass
concentration data for fermentation processes. The analysis is performed with only the on-line measurements of the
redox potential being available. It is a {\em two-step} procedure which includes an Artificial Neural Network
(ANN) that relates the redox potential to the biomass
concentrations in the first step. Next, a multifractal wavelet analysis is performed using the
biomass estimates of the process. In this context, our results show that the redox potential is a valuable
indicator of microorganism metabolic activity during the spontaneous fermentation.
In this paper, the detailed design of the multifractal wavelet analysis is presented, as
well as its direct experimental application at the laboratory level.

\end{abstract}


\end{frontmatter}

\section{Introduction}

During spontaneous fermentation processes, microorganisms employ sugars and other
constituents of must as substrate for their growth,
converting them into ethanol, carbon dioxide, higher alcohols and
their esters, and other metabolic compounds. In general, it could be useful to know more about the
dynamics of the entire microflora during spontaneous fermentations. In particular, an analysis that enables a monitoring
process could be fundamental for a quality control that ensures at
least the homogeneity of the final product.

Nevertheless, a bottleneck in all biochemical monitoring process is
often the lack of sensors for biological variables. Moreover, it is
a well-known issue that in order to monitor many biotechnological
processes, the problem of growth rate estimations represent a
strategic feature. Previously, various attempts of relating the \emph{redox potential}
to fermentation processes have been made taking into account that
\emph{redox potential} assesses the growth ability of
microorganisms, as well as the physiological activity in a given
environment \cite{Kwong 1992}, \cite{Berovic 1999}, \cite{van Dijk
2000}, \cite{Cheraiti 2005}. Particularly, the practical
significance of redox potential and oxygen content at various stages
of winemaking was examined in \cite{Kukec 2002}. Many chemical,
enzymatic and biological processes in wine are correlated with the
oxidative state of the wine.

It is well known that many natural systems exhibit complex dynamics described
by long-range power laws. In some cases the output of such systems, which corresponds to a
fluctuating time series, may be characterized or quantified by a spectrum of exponents called the
multifractal spectrum. The multifractal formalism to characterize processes, measures, and functions, introduced
by Halsey and collaborators~\cite{origen}, opened a new direction for the search of dimension type characteristics of dynamical systems
in the form of spectra for dimensions that reflect both the geometric and dynamical structure of nonlinear systems.
Despite the fact that there are a variety of methods to calculate the singularity spectrum of a multifractal structure,
the method based upon the wavelet theory appears to be the most appealing and successful \cite{arne0}.
In fact, the interest in wavelet methods is that they are numerically more stable~\cite{arne3}.
  In 1993, Bacry {\em et al.} \cite{arne1} developed this method based on the definition of a partition function in
which the concept of wavelet transform modulus maxima ({\sc WTMM}) was implemented. They demonstrated that the singularity
spectrum for a Bernoulli measure or a fractal distribution can be readily determined from the scaling behavior of such a
partition function and similar results have been proved for more general measures by Murgu\'{\i}a and Ur\'{\i}as \cite{Salo1}.

The paper is organized as follows. Section \ref{Smat} is devoted to a
concise presentation of the fermentation experiment performed to
illustrate our approach. The data analysis, which is a combination
of an ANN and a multifractal wavelet scheme, is described in Section
\ref{Ssoft}. Finally, the paper ends up with some
concluding remarks.

%>>>>>>>>>>>>>>>>>>>>>>>>>>>>>>>>>>>>>>>>>
\section{Experimental setup}\label{Smat}

\subsection{ Microorganism and culture conditions}

 In order to evaluate experimentally how suitable the estimation algorithm is,
 we performed six individual batch experiments with the must of \emph{Agave} (or \emph{Agave} syrup) for which we used inocula of
 native microorganisms (without the addition of any
commercial strain).
 This must was centrifuged at  $8000 \ rpm \times 10 \ min$ and stored
 in a frozen state at $-20 ^{\circ}C$ prior to experiments.

\begin{figure}[h]
\centering
          \includegraphics[height=8cm]{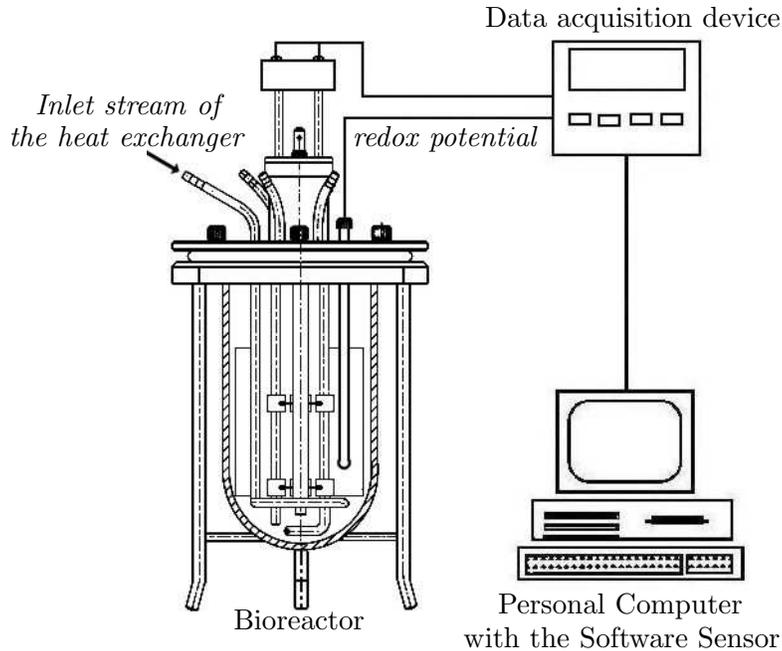}
            \put(-100,225){\footnotesize{Data acquisition device}}
            \put(-150,180){\footnotesize{\emph{redox potential}}}
            \put(-270,192){\footnotesize{\emph{Inlet stream of }}}
            \put(-280,180){\footnotesize{\emph{the heat exchanger}}}
            \put(-95,3){\footnotesize{Personal Computer}}
            \put(-108,-10){\footnotesize{with the Software Sensor}}
            \put(-195,-3){\footnotesize{Bioreactor}}
    \caption{The schematic representation of the experiments carried out in our laboratory.}
    \label{FigBioreactor}
\end{figure}

 The batch fermentations were carried out in a bioreactor  with pH and  redox sterilizable electrodes. The
 electrodes are connected to a console for data acquisition, a device which is connected to a computer where the data
 are stored and computed as can be seen in Fig.~\ref{FigBioreactor}.

 The bioreactor was filled with $900 \ mL$ of must as a culture medium,
 $100 \ mL$ of the inoculum in its exponential growing phase (biomass $0.1 \ g/L$)
  and $0.1$ \% of ammonium sulfate at final concentration.
The initial conditions of the fermentation were settled at a
temperature of $32.5 ^{\circ} C$ and initial sugar concentration of
$70 \ g/L$. The pH does not show a dynamic evolution, maintaining
itself at a value of $4$ during the whole process.
%For a schematic
%representation of the data acquisition process see
%Fig.~\ref{FigBioreactor}.

\subsection{Analytical procedures and measurements}

 The batch processes have been monitored for $14$ hours, through sampling under sterile conditions. In order
to quantify biomass and ethanol concentrations, $5 \ mL$ samples
of culture were removed every $30\ min$. The samples were cleared
by centrifugation at $6000\ rpm$ for $5$ minutes at room
temperature. The next step was to collect the supernatant phase
and store it frozen at $-20 ^{\circ}C$ prior to be analyzed. The
obtained pellet was resuspended in distilled water in order to
proceed with the biomass analysis.

The biomass measurements have been performed using (varian) UV
spectroscopy at $600 \ nm$. The obtained values were interpolated
with a standard curve of cell dry weight concentration.

%................ FIG 1

\subsection{Redox potential}

The measurement of \emph{redox potential} is relatively fast,
accurate, and reliable and its values give an insight into the
oxidation process as well as the inherent ability of reduction in
the process, which is well established in the case of wine
\cite{Kukec 2002}. The measured values of the \emph{redox potential} can give
information on redox reactions in wine, which have an important
effect on its quality and stability \cite{Kukec 2002}. During
storing and aging of wine, oxidation and reduction processes affect
the character and taste of wine to a considerable extent \cite{Kukec
2002}, \cite{Cheraiti 2005}. In our {\em Agave} case, the \emph{redox potential} measurements were acquired periodically
each $0.01\ hr$ during $14 \ hrs$, and the data were stored and
processed on line by means of a computer (see Fig.~\ref{FigBioreactor}).

\section{Data analysis} \label{Ssoft}

\subsection{Artificial neural network}

In order to relate the redox measurements to the
biomass concentration an ANN procedure is applied. The methodology
that we carried out includes a forward-propagation training
algorithm for the ANN using some of our experimental data. In
order to perform our task we construct a model of the following
form:
%...................
\begin{eqnarray}
    X_{1}=g(X_{2})
\end{eqnarray}
%.................
 where $X_{2}$ represents the \emph{redox potential}
measurement data ($mV$), $X_{1}=F(t)$ is the set of time-dependent
biomass concentration
data ($mg/L$) and the function $g(X_{2})$ is approximated by means
of the ANN procedure. The ANN architecture is of the standard type
\cite{lapedes} with a single ANN hidden-layer containing 10 units.

Each unit of this network uses a sigmoid function as the activation
function. On the other hand, the output contains a linear activation
function, in our case the identity. The feed-forward training
algorithm considered here is the conjugate gradient method
\cite{rumel}. Three of the six individual batch experiments were
used to provide data for the training process. The ANN after the
training gives an error of only 0.0029.

\subsection{Multifractal wavelet analysis}
%========================================

  The \emph{wavelet transform} (WT)
  of a distribution function $s(t)$ is given by %\in L^2(\mathbb{R})$ is given by
%...............
  \begin{equation}\label{eq2}
    W_s(a, b) = \frac{1}{a} \int_{-\infty}^{\infty} s(t)\bar{\psi} \left(\frac{t - b}{a} \right) dt,
      \label{eq-CWT}
 \end{equation}
  where $\psi$ is the analyzing wavelet, $b \in \mathbb{R}$ is a translation parameter, whereas
  $a \in \mathbb{R}^{+} ~ (a \neq 0)$ is a dilation or scale parameter, and
  the bar symbol denotes complex conjugation.

  One fundamental property that we require in order to analyze singular behavior is that
  $\psi(t)$ has enough vanishing moments~\cite{arne1, mallat}.
  A wavelet is said to have $n$ vanishing moments
  if and only if it satisfies

%%%%% ===========
\begin{equation} \label{momentosk}
  \int_{-\infty}^{\infty} t^k \psi(t) dt  = 0, \qquad \quad
  {\rm for} ~ k = 0, 1, \ldots , n - 1,
\end{equation}
and
\begin{equation} \label{momentoskn}
  \int_{-\infty}^{\infty} t^k \psi(t) dt  \neq 0, \qquad
  {\rm for} ~ k = n. %\notag
\end{equation}
%%%%% ===========
%%%%% ===========

  This means that a wavelet with $n$ vanishing moments is orthogonal to all
  polynomials up to order $n-1$. So the wavelet
  transform of $s(t)$ with a wavelet $\psi(t)$ with $n$ vanishing moments
  is nothing but a ``smoothed version'' of the $n$th derivative of
  $s(t)$ on various scales.

   In the presence of a singularity in the data at a particular
  point $t_0$, the scaling behavior of the
  wavelet coefficients is described by the H\"older exponent
  $\alpha(t_0)$ as

%%%%% ===========
\begin{equation} \label{eq-singular}
  W_s(a, t_0) \sim a^{\alpha(t_0)}~, %\qquad \text{in the limit $a \to 0^+$.}
\end{equation}
%%%%% ===========
%%%%% ===========
  in the limit $a \to 0^+$. %In the case of non-isolated singularities
%  case, is much better to consider the wavelet transform modulus maxima
%  (WTMM) \cite{mallat}. The singular behavior of signals
%  The location of the singularity can be detected, and
%  the related exponent can be recovered from the scaling of the WT,
%  along the so-called {\em maxima line}(WTMML) , converging towards
%  the singularity. This is a line where the WT reaches local maximum (with respect
%  to the position coordinate). Connecting such local maxima within the continuous
%  wavelet transform 'landscape' gives rise to the entire tree of maxima lines.
%  Restricting oneself to the collection of such maxima lines provides a particularly
%  useful representation of the entire WT. It incorporates the main characteristics
%  of the WT, the ability to reveal the {\em hierarchy} of (singular) features,
%  including the scaling behavior.

  The continuous wavelet transform given in Eq. \ref{eq-CWT}
  is an extremely redundant representation, too
  costly for most practical applications. To characterize
  the singular behavior of functions, it is sufficient
  to consider the values and position of the WTMM \cite{mallat}. The latter
  is a point $(a_0, b_0)$
  on the scale-time plane, $(a,b)$, where
  $|W_s(a_0, b)|$ is locally maximum for $b$ in the
  neighborhood of $b_0$.
  These maxima are located along curves in the plane $(a,b)$.
  However, the relationship (\ref{eq-singular})
  in some cases is not appropriated to describe
  distribution functions with non-isolated singularities.
  The wavelet multifractal formalism may characterize
  fractal objects which cannot be completely
  described by a single fractal dimension.
  According to Bacry {\em et al.} \cite{arne1}, %it is possible to associate
  an ``optimal'' partition function ${\mathcal Z}_q(s, a)$ can be defined in terms of
  the WTMM. %wavelet transform modulus maxima.
  They consider
  the set of modulus maxima at a scale $a$ as a covering of the singular
  support of $s$ with wavelets of scale $a$. The partition function
  ${\mathcal Z}_q$ measures the sum of all wavelet modulus maxima at a power $q$ as follows

%%%%% ===========
\begin{equation} \label{Mallat:FP}% \mathsf{Z}(a, q) = \sum_p |T_\psi[f](a, b_p(a)) |^q,
 {\mathcal Z}_q(s, a) = \sum_p |W_s(a, b_p(a)) |^q,
\end{equation}
%%%%% ===========

  where $\{ b_p(a) \}_{p \in \mathbb{Z}}$ is the position of all local maxima
  of $|W_s(a, b) |$ at a fixed scale $a$. This partition function
  is very close to the definition of the partition function described
  in \cite{origen}.
  It can be inferred from (\ref{Mallat:FP})
  that for $q>0$ the most pronounced modulus maxima will prevail,
  whereas for $q<0$ the lower ones will survive.
  For each $q \in \mathbb{R}$, the partition function %${\mathcal Z}_q(f, a)$
  is related to the scaling exponent $\tau(q)$
  %that characterizes its power--law behavior at fine scales $a$
  in the following way
  ${\mathcal Z}_q(s, a) \sim a^{\tau(q)}$.
  A linear behavior of $\tau(q)$ indicates monofractality
  whereas nonlinear behavior indicates a multifractal signal.
  A fundamental result in the (wavelet) multifractal formalism
  states that the singularity (H\"older) spectrum $f(\alpha)$ of
  the distribution function $s(t)$
  is the Legendre transform of $\tau(q)$, i.e.,
%%%%% ===========
\begin{equation} \label{Leg:D-tau}
     \alpha = \tau'(q), \qquad {\rm and} \qquad  f(\alpha)  = q\alpha - \tau(q).
\end{equation}
%%%%% ===========
The H\"older spectrum of dimensions $f(\alpha)$ is a
  nonnegative convex function that is supported on
  the closed interval $[\alpha_{{\rm min}}, \alpha_{{\rm max}}]$,
  which is interpreted as the Hausdorff fractal dimension of
  the subset of data characterized by the H\"older exponent $\alpha$ \cite{Salo1}.
  The most ``frequent'' singularity, which corresponds to the maximum  of $f(\alpha)$, occurs
  at the value of $\alpha(q=0)$, whereas the boundary
  values of the support, $\alpha_{{\rm min}}$ for $q>0$
  and $\alpha_{{\rm max}}$ for $q<0$, correspond to the
  strongest and weakest singularity, respectively.

 % The spectrum for dimensions $D(\alpha)$ and $\tau(q)$ are
%  smooth and strictly convex and they form a Legendre transform pair.
%  This allows one to compute the, otherwise intractable numerically,
%  spectrum for dimensions $f_\mu(\alpha)$ through the Hentshel Proccacia--spectrum.

  The analyzing wavelets which are used most frequently are the successive derivatives of the
  Gaussian function

   \begin{equation}\label{eq-Wavelets}
       \psi^{(n)}(t) := \frac{d^n}{dt^n} \left(\exp(-t^2 / 2)\right), \qquad n \in
       \mathbb{Z}^+, %=(-1)^nH_n(t)e^{-t^2}
   \end{equation}
because they are well localized both in space and frequency, and they remove the signal trends
that can be approximated by polynomials up to the $(n- 1)$th order \cite{arne3}. In particular, our analysis was carried out with the Mexican hat wavelet $\psi^{(2)}(t)$.

\section{Results and discussion}\label{Sres}

\begin{figure}[h!]
  %\hfill
 % \begin{minipage}[t]{.50\textwidth}
    \begin{center}
      \includegraphics[width=10cm,height=10cm]{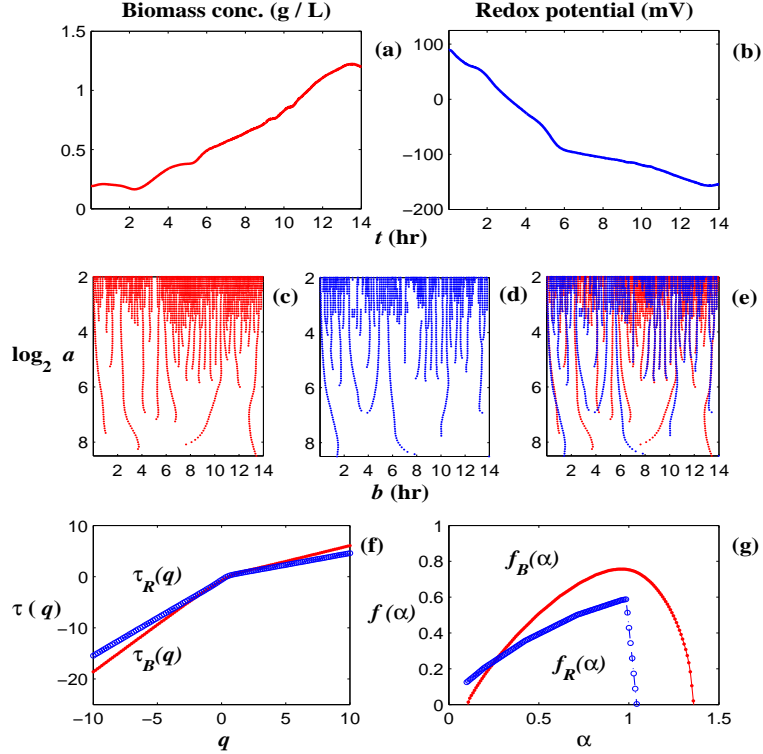}
      %\put(-200,195){\rotatebox{90}{\tiny{Biomass  (g/L)}}}
    \end{center}
     \hfill
      \caption{\small (a) Biomass signal for one of the six performed spontaneous fermentation processes of the {\em Agave} syrup (under the same experimental conditions) which was obtained by rebuilding from the redox potential data through the ANN procedure. The redox signal is displayed in (b). The maxima lines of the continuous wavelet transform of the biomass signal are shown in (c), of the transformed redox signal in (d), and in (e) one can see a comparison of their locations.
   (f) The corresponding $\tau (q)$ spectrum. (g) The multifractal singularity spectrum $f(\alpha)$ of the biomass and redox signals of this {\em Agave} spontaneous fermentation process.}\label{results1}
 % \end{minipage}
 \end{figure}

We performed the multifractal wavelet analysis based on the WTMM method employing the
biomass ANN estimates for two {\em Agave} spontaneous fermentation processes.
For the first one, we present the reconstructed biomass signal and the redox experimental signal in Fig.~(\ref{results1}) (a,b), respectively.
In general, for extracting the singularity structure of a signal according to wavelet transform techniques one should get the modulus maxima lines of the (continuous) wavelet transform. These are displayed in Fig.~(\ref{results1}) (c,d) and one can immediately notice that for both reconstructed and real experimental redox data we do not obtain isolated singularities as those treated in our previous work \cite{Ondeleta}, on the contrary there is a high density of singularities. This fact induces us to think of the multifractal formalism which could be more appropriate in such cases. The complete evidence of the multifractal character of the signals is a scaling exponent $\tau$ with two different slopes, where the point of bending in the $\tau (q)$ spectrum is directly related to the most frequent singularity of the biomass data. It is worth noting that
the locations of the maxima lines are almost the same for both the reconstructed biomass data and the direct experimental redox potential data, see the same figure (e). This  indicates that the effect of the ANN procedure on the singularity structure of the redox signal is small and that we can be confident in the ANN-based total biomass estimates. Thus, assuming that these spontaneous phenomena are multifractal processes, we calculated in
Fig.~(\ref{results1}) (f,g) the two basic multifractal quantities, namely the $\tau(q)$ and  $f(\alpha)$ spectra for both sets of data.
We confirm the multifractality of both signals since we indeed get a $\tau$ spectrum with two slopes in both cases.
The support of the singularity spectrum for the biomass signal is $\alpha \in [0.11, 1.36]$ with the maximum of the spectrum located at $\alpha _{peak}=0.95$ that gives us the most frequent singularity $f_{B}(\alpha _{peak})=0.756$, whereas in the case of the redox potential we get the support
$\alpha \in [0.1, 1.04]$ with $\alpha _{peak}=0.984$ giving the most frequent singularity $f_{R}(\alpha _{peak})=0.588$.
%We interpret the abrupt fall of the singularity spectrum in the case of the redox potential signal as due to the presence of a smooth component in %this signal similarly to the situation of a non-everywhere-singular fractal function as discussed by Muzy and collaborators \cite{arne0}.

\begin{figure}[h!]
    \begin{center}
      \includegraphics[width=10cm,height=10cm]{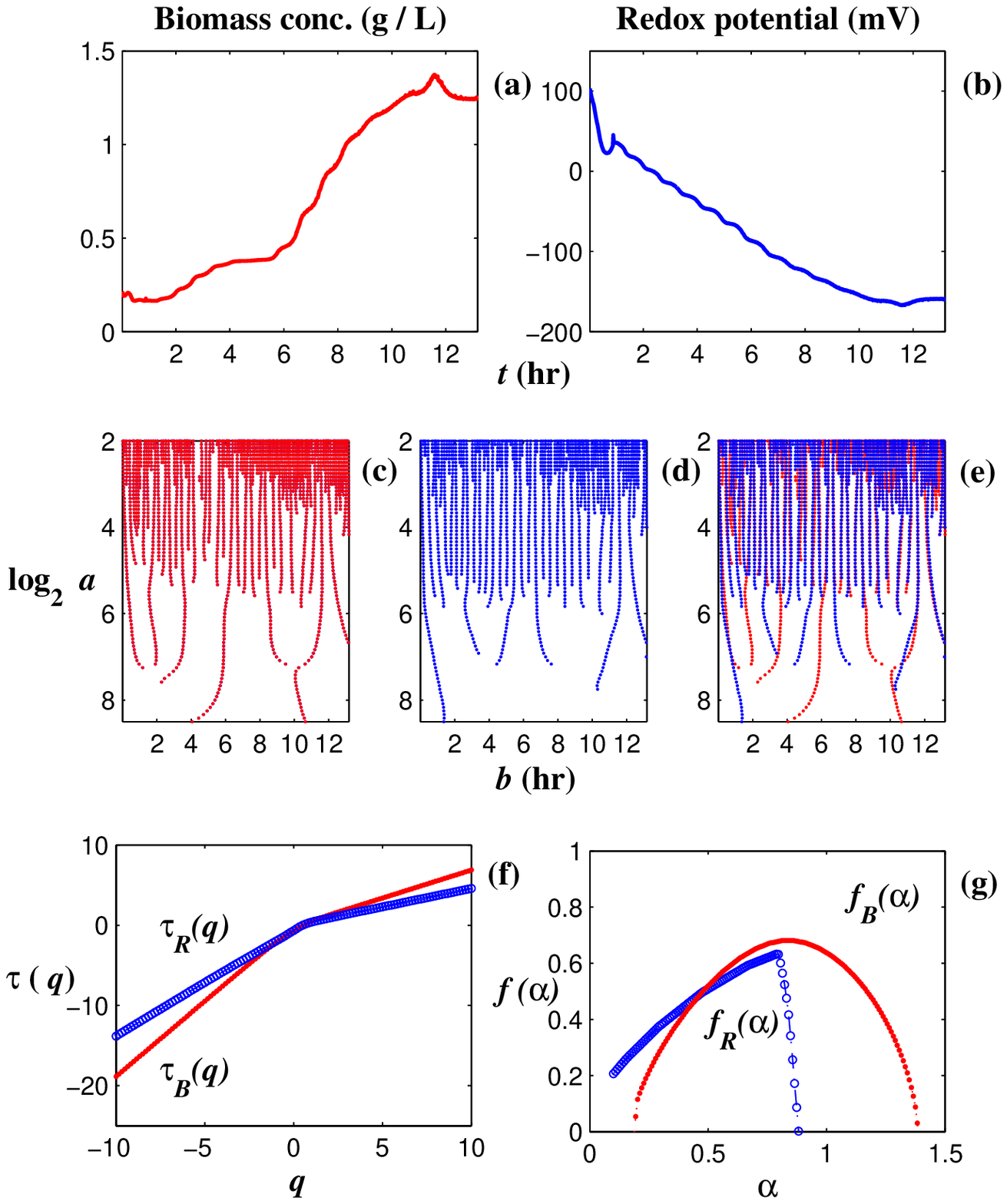}
       %\put(-200,195){\rotatebox{90}{\tiny{Biomass  (g/L)}}}
    \end{center}
    %  \end{minipage}
  \hfill
      \caption{\small (a) Biomass signal for another of the six spontaneous fermentation processes of the {\em Agave} syrup that we performed by rebuilding it through the ANN  from the redox potential data (same experimental conditions) presented in (b). The redox signal is displayed in (b). The maxima lines of the continuous wavelet transform of the biomass signal are shown in (c), of the transformed redox signal in (d), and in (e) one can see a comparison of their locations.
   (f) The corresponding $\tau (q)$ spectrum. (g) The multifractal singularity spectrum $f(\alpha)$ of the biomass and redox signals of this {\em Agave} spontaneous fermentation process.}\label{results2}
\end{figure}

We repeated the same multifractal analysis for another data set of the {\em Agave} spontaneous fermentation process to examine the reproduction of our results of the previous case. The second analysis is displayed in Fig.~(\ref{results2}). We again get the confirmation of the multifractal behaviour
from the two-sloped $\tau$ spectrum, but of course as we deal with a different time series we have slightly different multifractal parameters.
The support of the singularity spectrum for the biomass signal is now $\alpha \in [0.188, 1.38]$ with the maximum of the spectrum located at $\alpha _{peak}=0.84$ providing as the most frequent singularity $f_{B}(\alpha _{peak})=0.68$, whereas in the case of the redox potential we get now the support $\alpha \in [0.1, 0.88]$ with $\alpha _{peak}=0.7885$ that gives for the most frequent singularity the value $f_{R}(\alpha _{peak})=0.633$.

There is still one point to discuss, which is why the H\"older support extends up to one for the experimental redox potential signal whereas it goes
substantially beyond one in the case of the biomass signal in the two cases. We have found in the literature that a similar situation has been encountered by Humeau and collaborators \cite{hum} in the case of the laser Doppler flowmetry signals. The H\"older exponents of their simulated data are greater than one as compared to their real signals for which they are less than one. They comment that in general the simulated data are smoother whereas the real signals are much more irregular.

%This is related to the change of slope in the $\tau$ spectrum at the shoulder. % at In particular, in this situation the nonanalyticity of

%%.....................FIG 2
%\begin{figure}[h]
%  \hfill
%  \begin{minipage}[t]{.50\textwidth}
%    \begin{center}
%      \includegraphics[height=9cm]{CasoD1.eps}
%      %\put(-200,195){\rotatebox{90}{\tiny{Biomass  (g/L)}}}
%    \end{center}
%  \end{minipage}
%  \hfill
%  \begin{minipage}[t]{.45\textwidth}
%    \begin{center}
%      \includegraphics[height=9cm]{CasoD3.eps}
%       %\put(-200,195){\rotatebox{90}{\tiny{Biomass  (g/L)}}}
%    \end{center}
%      \end{minipage}
%  \hfill
%      \caption{(a) Biomass signal for two of the six spontaneous fermentation processes of the Agave syrup that we performed as rebuilt by the ANN  from the redox potential data (same experimental conditions). (b) The continuous wavelet transform of the two spontaneous fermentation biomass data sets.
%   (c) The Legendre transform member $\tau (q)$. (d) The multifractal singularity spectra $f(\alpha)$ of the two {\em Agave} spontaneous fermentations.}\label{results}
%\end{figure}

%.....................FIG 2
 % \hfill
  %\begin{minipage}[t]{.45\textwidth}

\section{Conclusions}\label{Sconc}

Although mixtures of microorganisms may look natural
in fermentation processes, the total
biomass, which is the quantity most easy to deal with, does not provide definite evidence for the mixture by itself.
However, in such processes the redox potential and the total biomass concentration are related fluctuating quantities thence
the redox potential signal may contain relevant information on the
microorganism metabolisms. This can be inferred through the multifractal techniques which has been specially designed for signals with significant irregular contributions.
%We attribute the H\"older exponents bigger than one that occur in the singularity spectra to a change in the carbon source of the microorganisms.

The main finding of this paper is that the spontaneous fermentation processes in the case of the {\em Agave} syrup can be characterized as a multifractal process and that the H\"older exponents are a good measure of the degree of irregularity of the signals. Since the total biomass concentration and the redox potential signals are connected spontaneous time series we expect to get quite similar singularity spectra for them, which indeed is the case. A detailed mapping of these H\"older irregularities to the metabolic parameters awaits more experimental data processed through the multifractal procedure.
In the literature one can find that many other natural processes have been approached by the multifractal means, such as the membrane ion activity of $K_{Ca}$ channels, i.e., their non-stationary dwell time series \cite{kaz}, the atmospheric energy cascades \cite{lovejoy}, and the point rainfall time series \cite{bend}, to name just a few. In the well-established multifractal techniques one can find embedded a useful amount of information and we strongly believe that the wavelet multifractal computational scheme may provide an
appropriate tool for quality control for this kind of spontaneous processes.
The results of this paper together with our previous research on mixed growth \cite{Ondeleta} show that the wavelet methods are a very useful tool for detecting physical singularities in the fermentation processes.

 \subsection*{Acknowledgements}

 VIJ received partial financial support from PROMEP, JSM received partial financial support from PROMEP and FAI-UASLP.
 HCR has been partially supported through the CONACyT project 46980.

\end{document}